\documentclass[12pt]{article}
\usepackage[utf8]{inputenc}

\usepackage{amsthm}
\usepackage{amsmath}
\usepackage{amssymb}
\usepackage[colorlinks,citecolor=blue,urlcolor=blue,filecolor=blue,backref=page]{hyperref}
\usepackage{graphicx}
\usepackage[authoryear, longnamesfirst, sort&compress]{natbib}
\shortcites{flaxman2020b, thompson2019, Nouvellet2018, verity2020, cintron-Arias2009, ferguson2006, gostic2020}

\newcommand{\ac}{\color{red} }  
\newcommand{\ca}{\color{black}} 
\newcommand{\cutac}[1]{\ac [cut] \ca} 

\newcommand{\mc}{\color{cyan} } 
\newcommand{\cm}{\color{black}} 
\newcommand{\cutmc}[1]{\mc[cut] \cm} 

\newcommand{\cg}{\color{blue} } 
\newcommand{\gc}{\color{black}} 
\newcommand{\cut}[1]{\cg[cut] \gc} 

\title{\Large \bf Filtering and improved 
Uncertainty Quantification in the dynamic estimation of effective  reproduction numbers}

\author{
Marcos A. Capistr\'an\footnotemark[1] \and
Antonio Capella\footnotemark[2] \and
J. Andr\'es Christen\footnotemark[1]\ \footnotemark[3]
}

\footnotetext[1]{Centro de Investigaci\'on en Matem\'aticas (CIMAT), Jalisco S/N, Valenciana, Guanajuato, GTO, 36023, MEXICO.
\textit{marcos at cimat.mx, jac at cimat.mx}}
\footnotetext[2]{Instituto de Matem\'aticas, Universidad Nacional Autónoma de México (UNAM).
\textit{capella at im.unam.mx}}
\footnotetext[3]{Corresponding author.}

\date{03DIC2020}

\begin{document}


\maketitle

\begin{abstract}
The effective reproduction number $R_t$ measures an infectious disease's transmissibility as the number of secondary infections in one reproduction time in a population having both susceptible and non-susceptible hosts. Current approaches do not quantify the uncertainty correctly in estimating $R_t$, as expected by the observed variability in contagion patterns. We elaborate on the Bayesian estimation of $R_t$ by improving on the Poisson sampling model of \cite{cori2013}. By adding an autoregressive latent process, we build a Dynamic Linear Model on the log of observed $R_t$s, resulting in a filtering type Bayesian inference. We use a conjugate analysis, and all calculations are explicit. Results show an improved uncertainty quantification on the estimation of $R_t$'s, with a reliable method that could safely be used by non-experts and within other forecasting systems. We illustrate our approach with recent data from the current COVID19 epidemic in Mexico.
\end{abstract}

\newpage

\section{Introduction}

The effective reproduction number $R_t$ is the proportion of new cases generated by active cases at calendar time $t$.  Estimating $R_t$ has proved to be an important tool for monitoring ongoing epidemics, to monitor interventions or general changes in contagion patterns \citep[e.g.][]{ferguson2006, fraser2004}.  This has been the case in the past and even more recently with the current COVID19 epidemics.  In Germany, a calculation of $R_t$ is used as public monitor of the evolution of the COVID19 epidemic and has direct consequences on public life and decision making \cite{RKI2020}.  Regarding the estimation of $R_t$, in basically all realistic cases, a proxy for incidence data has to be used, and along with the inherent stochastic variability in reported cases, proxies are subject to noise. 

There are methods to estimate $R_t$ as a sub product of larger mechanistic compartmental models.   The main limitation of these methods is precisely that an underlying mechanistic model must be postulated and calibrated beforehand, thus inheriting possible model drawbacks \citep[see][for example]{bettencourt2008, cintron-Arias2009}.  Recently, more generic methods for estimating $R_t$, requiring incidence data only, have been proposed \cite{cauchemez2006, wallinga2004, fraser2007, cori2013, thompson2019}.  \cite{cauchemez2006} propose a hierarchical model and Bayesian estimation using Markov Chain Monte Carlo (MCMC) methods to estimate $R_t$.  MCMC is hardly recommended for non-experts and simpler analytic methods are sough.  

Let $I_t$ be the incidence of cases, to be used for the estimation of $R_t$, $t = 1, 2, \ldots$.
Using the renewal equation and other assumptions \cite{fraser2007} explains that
the effective reproduction number $R_t$ may be defined and estimated with
\begin{equation}\label{eqn:Rt}
R_t = \frac{I_t}{\sum_{s=1}^t I_{t-s} w_s}  = \frac{I_t}{\Lambda_t}.
\end{equation}
Here $w_s$ is a probability mass function (i.e. $w_s \geq 0$ and $\sum_{s=1}^\infty w_s = 1$) that accounts for the infectious disease generation interval.
It is assumed that infected individuals have a generation interval $w_s$, dependent on the number on days since infection $s$ but independent of current time.  At time $t$, $I_{t-s} w_s$ represents a ``force of infection'' of individuals infected $s$ days ago.  Then $\Lambda_t$ represents the ``total infectiousness of infected
individuals'' \cite{cori2013}.  It may also be seen as an estimation of the current total of active cases.  $R_t$ is the ratio of secondary cases $I_t$ produced by the current total of active cases. From here onwards we will assume that time $t$ is measured in days.

Note that $I_t$ may be only a proxy of the total number of infected people, as for example, people arriving at hospitals to seek help, etc.  In most common cases, the new infected individuals at time $t$, $I_t^*$, including asymptomatic infections, is impossible to measure.  Ideally, as a proxy, we may regard instead $I_t = K I_t^*$ with $K$ an unknown constant.
However, $R_t$ is a unit less measure of the proportion of secondary infections and is indeed independent of $K$.
Any estimation of $R_t$ should reflect this, providing the same results if a different $K$ is used or if the proxy is given proportional to the total population etc.  Moreover, the uncertainty we have on the actual $R_t$ should not be bound to the absolute values of $I_t$ but to the recent variation in the observed ratio $I_t/\Lambda_t$.

Moreover, epidemic data has far more caveats that need to be address.  Reporting delays , along with right truncation are very common in epidemic data.  Only under a thoughg investigation of the reporting system at hand, is that incidence data may be corrected to obtain a reasonable enough good proxy for estimating $R_t$ \citep{salmon2016, mcgough2020}. As we say, only ideally is that
$I_t = K I_t^*$.  Even after standard data pre processing techniques are applied, conditions may change during an epidemic with the net result that $K$ changes over time, for example.  Undoubtedly, the use of raw incidence data, without proper pre processing, most likely will lead to incorrect/biased estimates of $R_t$ \citep{gostic2020}.
We do not discuss data preprocessing any further and assume $I_t = K I_t^*$ throughout, with an unknown fixed $K$.

In \cite{cori2013} the estimation of $R_t$, based on the proxy incidence $I_t$, uses (\ref{eqn:Rt}) to form the statement that
$$
E(I_t | I_{t-1}, I_{t-2}, \ldots, I_1 ) = R_t \Lambda_t .
$$
A probability model is then proposed for the observed $I_t$ conditional on $I_{t-1}, I_{t-2}, \ldots, I_1$ with the expected value
$R_t \Lambda_t$.  \cite{cori2013} proposed the Poisson model $I_t | I_{t-1}, I_{t-2}, \ldots, I_1 \sim Po(R_t \Lambda_t)$.  Incidence data is commonly overdisperse and the immediate improvement on this Poisson model would be to use a more general count data model as a Negative Binomial \citep{linden2011}.  Dispersion parameters may be fixed or estimated in the Negative Binomial \citep{flaxman2020b}.  However, directly postulating a model for incidence data may result in estimates that depend on the unknown constant $K$.  Moreover, as far as estimating $R_t$ is concern, the actual variability of $I_t$ is not in principle relevant, only the variability observed in the ratios  $I_t/\Lambda_t$.

\begin{figure}
\begin{center}
\includegraphics[width=13cm,height=5.5cm]{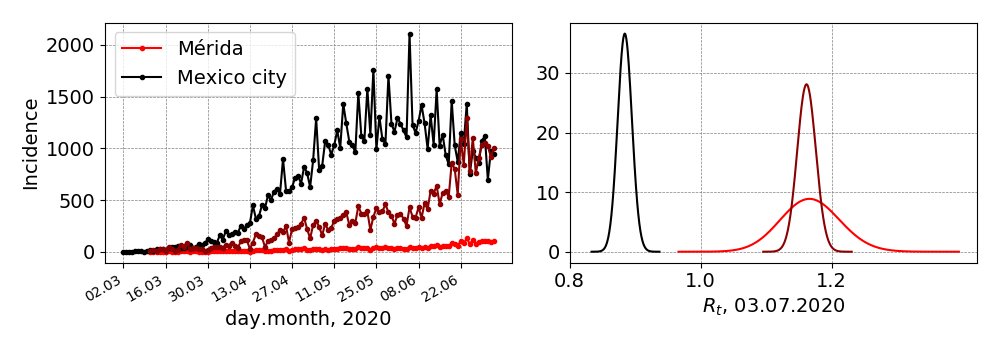} \\
\caption{\label{fig:compare} Incidence data for the ongoing COVID19 epidemic in two metropolitan areas of Mexico
\citep[left, confirmed cases, see][]{capistran2020}.  The corresponding posterior of $R_t$ according to \cite{cori2013} Poisson model (right).  Also, the Mérida incidence data multiplied by 10, and its corresponding posterior (dark red).  The uncertainty presented by the posterior responds to the incidence absolute values and not to the variation in reproduction numbers.  The generation interval $w_s$ used here is explained in Figure~\ref{fig:compare_sn}.}
\end{center}
\end{figure}

The resulting inference of \cite{cori2013}, where they assume a constant $R_t^\tau$ of $\tau=7$ days (with a gamma
$Ga( 5, 1)$ conjugate prior) is illustrated in Figure~\ref{fig:compare}.  In particular, the coefficient of variation (CV) of the posterior distribution for $R_t^\tau$ is
$$
\frac{1}{\sqrt{a + \sum_{t-\tau +1}^t I_t}} 
$$
\citep[see][p.3 web material]{cori2013}.  The CV will decrease if we increase the arbitrary constant $K$.  It does depend only on the absolute current values for $I_t$ and not on the intrinsic variation of the underlying $R_t$'s, see Figure \ref{eqn:Rt}.  Further elaboration on this Poisson model may be found in recent literature.
\cite{thompson2019} improves this Bayesian inference of $R_t$ by also estimating $w_s$ and adding the corresponding variability in the posterior of $R_t$.  
\cite{Nouvellet2018} uses \cite{cori2013} embedded in a more complex forecast system.
 The need for a statistical estimation of $R_t$ is apparent, with a reasonable estimation of the uncertainty in all inferences.

Recently, and prompted by the current uses of $R_t$ estimates during the COVID19 epidemic, \cite{gostic2020} make a review of current methods and recommend the use of \cite{cori2013}, in contrast to other methods not based on
(\ref{eqn:Rt}), see \cite{gostic2020} and references therein.  \cite{gostic2020} stress some other three points in the estimation of $R_t$, namely: 1) Data for estimating the infectivity function $w_s$ is rarely available, and proxies for estimating $w_s$ should be used with care.  2) Correct incidence data should always be used, representing the number of new infections every day, taking special care to correct for reporting delays and other issues. 3) Smoothing of $R_t$ estimation is a matter of concern.  \cite{cori2013} uses a smoothed window of $\tau=7$ days, but this should be used judiciously depending on the data at hand, although no definite guidelines are suggested in \cite{gostic2020}.  However, \cite{gostic2020} focus on other comparing issues in the estimation of $R_t$ and did not concentrate on the adequateness of the uncertainty expressed by the posterior distributions of \cite{cori2013}.      


Our aim here is to improve the Uncertainty Quantification (UQ) in the estimation of $R_t$, from the definition in (\ref{eqn:Rt}), and using an analytic Bayesian approach (i.e. non-MCMC) that may be robustly embedded in other systems or used by non-experts. 

We follow the same basic idea of \cite{cori2013} but interpret (\ref{eqn:Rt}) differently.  Namely, define
$\rho_t = log(R_t)$, then
$$
E(i_t | i_{t-1}, i_{t-2}, \ldots , i_1, \rho_t, \rho_{t-1}, \rho_{t-2}, \ldots , \rho_1 ) = \rho_t + \lambda_t ,
$$
where $i_k = log(I_k) + log(K)$ and $\lambda_t = log(\Lambda_t) + log(K)$, for some unknown constant $K$.  As usual, when taking logarithms of count data to postulate an observation model we assume that $i_t | \cdots ~ \sim N(\rho_t + \lambda_t, \sigma)$, for some unknown variance $\sigma$, or equivalently
$$
y_t = i_t - \lambda_t \sim N( \rho_t, \sigma_t^2).
$$ 
As recommended by \cite{cori2013}, $R_t$ should only be estimated for incidence with $I_t \geq 10$.   This further justifies the use of the above Normal model.
Therefore, with the observed log $R_t$'s, i.e. the $y_t$s, we will try to estimate the $\rho_t = log(R_t)$s. 
 \cite{cori2013} assumed a constant $R_t^{\tau}$ over $\tau$ days, obtaining some smoothing.   A further improvement on the approach of \cite{cori2013} is that we explicitly and formally model smoothing, or coherence, among the $\rho_t$s.  This we do by postulating an autoregressive prior for the $\rho_t$'s, modeling the fact that $\rho_t$ should be \textit{similar} (but not necessarily equal) to $\rho_{t-1}$. This creates a Dynamic Linear Model, and using Bayesian updating, forms a filtering type inference for the $\rho_t$ sequence.
 
 We will continue to assume the generation interval $w_s$ fixed and dedicate our effort to improving the UQ in the Bayesian estimation of $R_t$.  As mentioned above, we do not discuss data preprocessing and assume that $I_t$ is a correct proxy for estimating $R_t$.  We do not provide an analysis on the effects of incidence data anomalies on our estimation of $R_t$.
The paper is organized as follows.  We present the details of our approach in the next section and in section \ref{sec:results} we present two examples using recent COVID19 incidence data from Mexico.  Finally, in Section \ref{sec:dis} we make a discussion of our results.

\section{Materials and methods}

\begin{table}
\caption{\label{tab:formulas} Updating formulas for the parameters of the non-central student-t posterior distribution for
$\rho_t$, $\rho_t | D_t \sim T_{n_t} ( m_t, c_t )$.}
\medskip
\begin{tabular}{| l l l |}\hline
& & \\
$n_t = \delta n_{t-1} + 1$ & $a_t = g m_{t-1}$ & $r_t^* = c_{t-1} + w^*$ \\
$q_t^* = ( r_t^* +1 )$ & $q_t = s_{t-1} q_t^* $ & $r_t = s_{t-1}  r_t^*$\\
$e_t = (y_t - a_t)$ & $s_t = \delta \frac{n_{t-1}}{n_t} s_{t-1} + \frac{s_{t-1}}{n_t} \frac{e_t^2}{q_t}$ &
$A_t = \frac{r_t^*}{q_t^*}$ \\
& & \\
$m_t = a_t + A_t e_t$ & $c_t = \frac{s_t}{s_{t-1}} \left\{ r_t - A_t^2 q_t \right\}$ &\\ 
& & \\ \hline
\end{tabular}
\end{table}

A Dynamic Linear Model \citep{west1997} is proposed for the estimation of the $\rho_t$'s.  Namely,
\begin{eqnarray*}\label{eqn:model}
y_t = \rho_t + \nu_t, & \nu_t \sim N( 0, \phi_t^{-1}). \\
\rho_t = \rho_{t-1} + \omega_t, & \omega_t \sim N( 0, \phi_t^{-1}  w^*). \\
\phi_t = \gamma_t \phi_{t-1} / \delta, & \gamma_t \sim Beta( \delta n_{t-1}/2, (1-\delta) n_{t-1}/2).
\end{eqnarray*}
With $\rho_0 \sim N( m_0,  \phi_t^{-1}  c_0^*)$ an autoregressive AR(1) (prior) model is proposed to estimate the log $R_t$'s.  Here we state that $\rho_t$ is equal to $\rho_{t-1}$ plus some noise $\omega_t$.
The precision $\phi_t$ process, i.e. $\sigma_t^2 = \phi_t^{-1}$, is the variance of the observed log $R_t$'s, $y_t$, and is assumed unknown.  With $\gamma_1 \sim Beta( \delta n_{0}/2, (1-\delta) n_{0}/2)$ a second AR(1) is assumed to estimate these variances.  $n_t$ are the discounted degrees of freedom $n_t = \delta n_{t-1} + 1$,
\citep[see][sec. 10.8 for details]{west1997}.  Here we also assume that the precision $\phi_t$ is similar to $\phi_{t-1}$ multiplied by the random innovation or shock $\gamma_t  / \delta$.  Since
$E(\gamma_t) = \delta$, $\gamma_t  / \delta$ has a shifted $Beta$ distribution with mean 1.  This keeps all $\phi_t$ positive. 
The \textit{discount factor} $\delta$ and the hyperparameters
$m_0, c_0^*, w^*, n_0$ are part of the modeling and prior definition (see below) and considered known.  

The marginal posterior of $\rho_t$ may be calculated analytically and corresponds to a non-central student-t distribution with $n_t$ degrees of freedom
\begin{equation}\label{eqn:rhot_post}
\rho_t | D_t \sim T_{n_t} ( m_t, c_t )
\end{equation}
where $D_t  = (y_t , D_{t-1})$; see Table \ref{tab:formulas} for the recursive calculation of the central parameter $m_t$ and dispersion parameter $c_t$.  This constitutes a filtering type estimation \citep{kalman1960} of the $\rho_t$'s, with discounted variance \citep[][sec. 10.8]{west1997}.  The marginal posterior distribution of $R_t$ may be obtained from
(\ref{eqn:rhot_post}), namely
$$
f_{R_t }( r | D_t ) = r^{-1} f_{\rho_t}( log(r) | D_t).
$$
In particular, quantiles for this posterior may be calculated by transforming the corresponding quantiles of (\ref{eqn:rhot_post}) since $R_t = exp(\rho_t)$ is monotonic.

\subsection{Autoregressive modeling and prior distributions}\label{sec:priors}

The assumption of \cite{cori2013} that $R_t$ remains constant over $\tau$ days is relaxed here with the AR(1) prior model
for the $\rho_t$'s, $\rho_t = \rho_{t-1} + \omega_t, \omega_t \sim N( 0, \phi_t^{-1} w^*)$.  
The variance discount factor $\delta$ is key for learning from recent variability in the $\rho_t$'s and gradually cease to learn from more distant  $\rho_t$'s.  It is proved that the degrees of freedom $n_t$, which we may visualize as the ``effective'' sample size, the number of sample points in the past actually used to estimate the variance component, has the property $n_t \rightarrow (1-\delta)^{-1}$ \citep[][sec. 10.8]{west1997}.  We postulate that $(1-\delta)^{-1} = 2 \tau$, or $\delta = 1 - (2 \tau)^{-1}$.  That is, the limit to learning from the past is $2 \tau$ days.
Regarding the prior for $\rho_0 \sim N( m_0, \phi_0^{-1} c_0^*)$, we center it at $m_0=0$ (i.e. $R_t = 1$) with the same variance for the observed $y_t$, that is $c_0^* = 1$.
Regarding the prior for $\gamma_1 \sim Beta( \delta n_{0}/2, (1-\delta) n_{0}/2)$, note that by design it is centered at $\delta$.  Using $n_0 = 2$ provides a diffuse prior with large variance, and leads to the default non informative $Beta( 1/2, 1/2)$ in the case of $\delta = 1/2$.

The variance for the AR(1) model for $\rho_t$ is crucial, as it controls the memory and smoothness in our approach.
Our a priori statement on this variance is the value of the variance multiplier $w^*$.
The prior model for $\rho_t$ is in fact a summation of Gaussian shocks (as in the Weinner process).  It is clear then that
$\rho_t | \rho_{t-k} \sim N( \rho_{t-k}, \phi_t^{-1} k w^*)$.  If the variance is similar to, or larger than, the variance of the observational process, that is $k w^* \geq 1$, then little smoothing is obtained at lag $k$ since $\rho_t$ can vary as much as, or more, than $y_{t-k}$.  Taking $k=\tau$, we set $\tau w^*= 2$, thus $w^*= 2/\tau$, to also limit the level of smoothing to less than $\tau$ days.  We postulate $\tau w^*= 2$ as a pragmatic large enough variance multiplier, that leads to a quite similar smoothing obtained by \cite{cori2013} using the same value for $\tau$.  This is illustrated in the examples presented in the next Section.

\section{Results} \label{sec:results}

\begin{figure}
\begin{center}
\begin{tabular}{c c}
\includegraphics[scale=0.5]{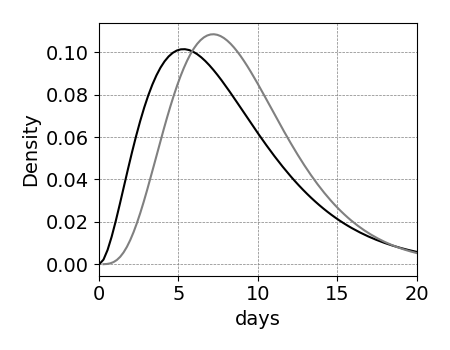} &
\includegraphics[scale=0.5]{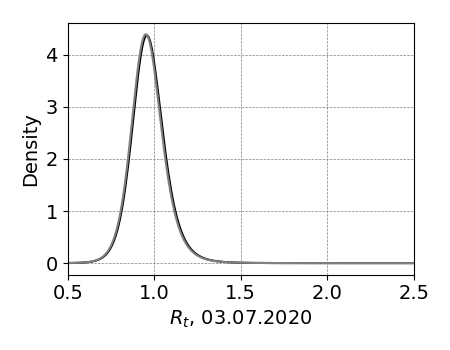} \\
(a) & (b)
\end{tabular}
\caption{\label{fig:compare_sn} (a) Our postulated Erlang( 3, 8/3) (black) and slightly shifted Erlang( 5, 9/5) (grey), as the generation interval of COVID19. (b) The resulting posterior distributions of $R_t$ on 03JUL2020 for our method, for the Mérida COVID19 incidence data presented in Figure~\ref{fig:compare}.}
\end{center}
\end{figure}

A key input to calculating $R_t$ is postulating  $w_s$ through an infectious disease generation interval.  $w_s$ may be seen as the probability ($\sum_{s=1}^\infty w_s = 1$) that an infected person infects other people $s$ days from infection \citep{fraser2007}.  A readily way to define $w_s$ is by defining a pdf $f(s)$ and with its cdf let $w_s = F(s) - F(s-1)$.  For the COVID19 epidemic, we postulate an Erlang( 3, 8/3) pdf depicted in Figure~\ref{fig:compare_sn}(a).    The expected value is at 8 days and the maximum infectivity is at 5.5 days, decaying near zero after 20 days.  This is based on early \citep{eurosurveillance2020, verity2020} as well as recent reports on the viral load and implied infectiousness of the decease \citep[see][fig. 2]{cevikm2020}.  The use of the Erlang distribution in epidemics has been suggested elsewhere \citep{champredon2018}.

To illustrate that our inferences are stable, we also present an Erlang generation interval shifted to the right, that may also be considered adequate for COVID19.  The resulting posterior distributions for the $R_t$ using our method, for Mérida, M\'exico, on 03JUL2020 are presented in Figure~\ref{fig:compare_sn}(b).  Certainly, under broad regular circumstances, the Bayesian posterior operator is smooth with respect to changes in the prior \citep[see][for example]{christen2020}.
We continue with an Erlang( 3, 8/3) as our generation interval distribution for COVID19 in both examples below.  

\begin{figure}
\begin{center}
\begin{tabular}{c}
\includegraphics[scale=0.7]{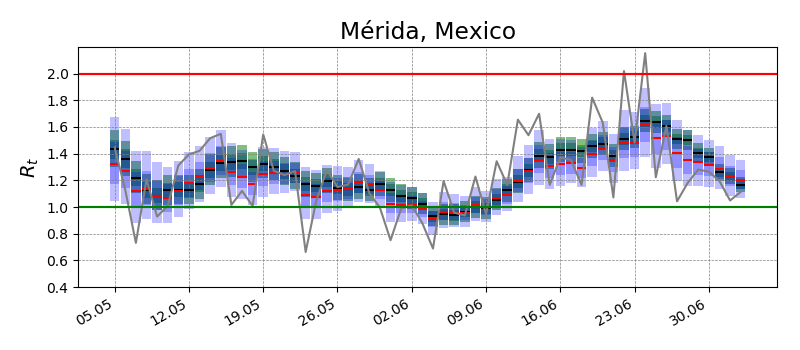}\\
\includegraphics[scale=0.7]{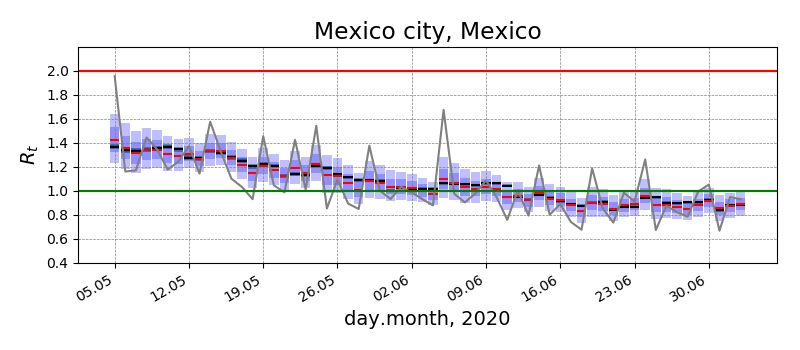}\\
\end{tabular}
\caption{\label{fig:results} $R_t$'s for incidence data presented in Figure~\ref{fig:compare} for the last 60 days. From the posterior marginal of each $\rho_t$ in (\ref{eqn:rhot_post}) we calculate its 10\%, 25\%, 50\%, 75\% and 90\% quantiles and these are transformed back to obtain quantiles of $R_t$.  These are plotted vertically with blue shades and the median with a red line.  The Poisson model posterior of $R_t$ of \cite{cori2013} is also plotted, using the same quantiles with green shades and a black line for the median. The posterior of \cite{cori2013} for Mexico city is so sharp that the quantile shades are hardly seen.  The gray lines are the observed $R_t = e^{y_t}$s.}
\end{center}
\end{figure}

We use the same data sets presented in Figure~\ref{fig:compare} to illustrate our approach, plotting the posterior $\rho_t$ time series from the last date in the time series to 60 days before, see Figure~\ref{fig:results}.
Also plotted in Figure~\ref{fig:results} are the posterior for the $R_t$s using \cite{cori2013} and the observed $R_t=e^{y_t}$s.  Regarding the Mérida data, note how \cite{cori2013} posterior is basically insensitive to the variability observed in the data, the posterior inter-quantile ranges remain basically equal.  For our approach, note from 05.05 to 19.05 the large variability in
$e^{y_t}$ leads to large posterior ranges whereas from 02.06 to 09.06 a more stable observed $R_t$ results in far shorter posterior inter-quantile ranges.  Later on, by the end of June, more spread posteriors are seen again given the observed second wave of large and unstable $R_t$s.  A sharp increase in $e^{y_t}$ follows a sharp decrease, from above 2 to nearly 1, within 10 days until 03.07.  Our AR(1) modeling results in a smooth decay on the position of the posterior for $R_t$, and shorter spread, as a response to the short stable spell of observed  $e^{y_t}$.  For these same 10 days, a similar behavior is seen in the position of  \cite{cori2013}'s posterior, but this as a result of the moving window of $\tau$ days, as explained in the previous section.
As already mentioned, the UQ resulting from the posterior of \cite{cori2013} responds only to the absolute values of $I_t$s, which in this case are around 25 to 100 in May and June.

Regarding the results from Mexico city, presented in Figure~\ref{fig:results}, note how the posterior of \cite{cori2013} basically collapses around the median, given the large $I_t$ counts of this large metropolitan area (with a population of more than 21 million, $I_t$ varies around 1000 to 2000 in May and June).  On the contrary, our UQ provided by the posterior of $R_t$s presents larger spreads in early May, when an unstable spell is observed for $e^{y_t}$.  Later on, after most of June with a more stable period, the uncertainty decreases with far sharper posteriors by the end of the month and the firsts days of July.

Regarding the smoothing obtained in the estimation of $R_t$ with respect to the observed values, that is, gray lines in Figure~\ref{fig:results}, see how sharp changes in the latter are reasonably smoothed out by the former.
Our strategy to setting the variance multiplier hyperparamenter $w^*$ in the AR(1) model for $\rho_t$ (i.e. $w^* = 2/\tau$, with $\tau=7$ days, see Section~\ref{sec:priors}) results in a smoothing similar to that of \cite{cori2013}'s, although with larger and adaptive inter-quantile ranges, as already commented.  The smoothness obtained by \cite{cori2013} is a result of assuming a fixed $R_t$ for $\tau=7$ days and calculating the posterior independently for each $R_t$.  Ours is the result of the DLM and the autoregressive prior model on $\rho_t$.

\section{Discussion}\label{sec:dis}

We present a new approach for estimating the effective reproduction number $R_t$ with the following improvements:  1) It does not depend on a compartmental model, specific to a particular disease. 2) Using previous ideas  \citep{fraser2007, cori2013} we construct a novel statistical model and Bayesian inference approach to estimate $R_t$. 3) The resulting UQ is better suited responding to the recent variation of observed $R_t$s. 4) An AR(1) process is used to promote smoothness on the estimation and 5) all calculations are simple, leading to a Bayesian updating (filtering) type analysis that may be used by non-experts.
In fact, a simple to use Python program with an implementation of our method \citep[and also the original of][]{cori2013} may be downloaded from \url{https://github.com/andreschristen/RTs}.  We are working also on an alternative implementation in a spreadsheet.

The generation interval $w_s$ is a crucial input to this and other similar methods to estimate $R_t$.
Note that based on preliminary data on infected patients, some basic knowledge can be gained to postulate a reasonable $w_s$.  This was the case since early stages of the COVID19 epidemic \citep{eurosurveillance2020}.  It is easier to gain information on  $w_s$ since it represents information on the infectivity, bound to an average infected person.  Social contact patters, lockdown measures etc do not play a role in $w_s$ \citep[from the underlying assumption that $\beta( t, s)$, in the renewal equation
$I(t) = \int   \beta( t, s) I(t-s) ds$, has the form $\beta( t, s) = R(t) w(s)$, see ][]{fraser2007}.  The social/epidemiological factors influencing contagion patters are coded in $R_t$, an appealing feature of this type of approaches to estimate the latter, to monitor changes in epidemic data from the onset.  


Estimating the infectious disease generation interval $w_s$ requires time consuming clinical studies that are commonly not available during an ongoing epidemic with a new virus as Sars-Cov-2.  \cite{thompson2019} present an attempt to include uncertainty in the estimation of $w_s$ into the posterior distribution of $R_t$ of \cite{cori2013}.  Including an estimate of $w_s$  in our method, along with its corresponding uncertainty, is formally straightforward within this Bayesian framework, either by including a likelihood with additional observables or by postulating a prior process for $w_s$.  Indeed, this will complicate the approach and most likely MCMC methods will need to be used.  

As already mentioned, a reliable estimation of $R_t$ is quite relevant, to be used as such, or as part of other larger forecasting or monitoring systems. \cite{bettencourt2008} elaborate on a simple relation of future incidence with predicted $R_t$s.  A reliable forecasting system of $R_t$ could then be used for short term forecasting of $I_t$. A related idea is used in \cite{Nouvellet2018} for incidence forecast and other our purposes.
Being at time $t$, the process $\rho_{t+k}, k= 1, 2, \ldots, p$ can be easily simulated, by simulating $\rho_t$ and other parameters from the posterior at $D_t$ and in turn simulating $\rho_{t+k} | \rho_{t+k-1}$ until $\rho_{t+p}$, to produce MC samples of the posterior predictive distribution for $R_t$.   A feasible path for incidence forecasting based on forecasts of $R_t$ is envisaged, but we need to leave this possibility for future research.

\section{Acknowledgments}

The authors are partially founded by CONACYT CB-2016-01-284451 and COVID19 312772 grants and a RDCOMM grant.
AC was partially supported by UNAM PAPPIT–IN106118 grant.

\bibliographystyle{chicago}
\bibliography{Rts_AR.bib}
\end{document}